\documentclass[referee]{raa}
\usepackage{times}
\usepackage{natbib}
\usepackage{epstopdf,cite,mathtools,ifthen,graphicx,xcolor,setspace}
\usepackage{textcomp,array,booktabs,epsfig,rotating,subfigure,setspace}
\usepackage{SIunits}
\bibpunct{(}{)}{;}{a}{}{,}

\usepackage[a4paper=true,dvipdfm=true,pagebackref=true]{hyperref}
\hypersetup{pdftitle = The title of my PDF, pdfauthor = My name, pdfsubject= The subject, pdfkeywords = keyword1 keyword2 keyword3}
\hypersetup{colorlinks = true, linkcolor = green, anchorcolor = red, citecolor = blue, filecolor = red, pagecolor = red, urlcolor = red}

\begin{document}

\title{Upgraded Photometric System of The 85-cm Telescope at Xinglong Station$^*$$^\dag$
\footnotetext{\small $*$ Supported by Key Laboratory of Optical Astronomy,
National Astronomical Observatories, Chinese Academy of Sciences.}
\footnotetext{\small $\dag$ Supported by the National Natural Science Foundation of China (NSFC) No. 11403088 , 11273051, 11673003 and the National Basic Research Program of China (973 Program 2014CB845700)}
}

 \volnopage{ {\bf 2017} Vol.\ {\bf X} No. {\bf XX}, 000--000}
   \setcounter{page}{1}

   \author{Chunhai Bai\inst{1,2,3}, Jianning Fu\inst{4}, Taoran Li\inst{1}, Zhou Fan\inst{1},  Jianghua Wu\inst{4}, Yong Zhao\inst{1}, Xianqun Zeng\inst{1}, Wenzhao Zhang\inst{4}, Peng Qiu\inst{1}, Guojie Feng\inst{3}, Xiaojun Jiang\inst{1}
   }

   \institute{ National Astronomical Observatories, Chinese Academy of Sciences,
20A Datun Road, Chaoyang District, Beijing 100012, China; {\it baichunhai@xao.ac.cn}\\
   \and  University of Chinese Academy of Sciences, Beijing 100049, China\\
   \and  Xinjiang Astronomical Observatories,  Chinese Academy of Sciences,
Scientific 1st street, Urumqi, Xinjiang 830011, China\\
    \and Department of Astronomy, Beijing Normal University, Beijing 1000875, China\\
\vs \no
   {\small Received 2017 May 30; accepted 2017 July 27}
}
\abstract{The 85-cm telescope at the Xinglong station is a well-operated prime focus system with high science outputs. The telescope has been upgraded since 2014 with new corrector, filters and camera, which are provided by Beijing Normal University (BNU). The filter set is Johnson-Cousins UBVRI system. We report the test results of the new system including the bias, dark current, linearity, gain and readout noise of the CCD camera . Then we derive accurate instrumental calibration coefficients in UBVRI bands with Landolt standard stars in the photometric nights. Finally, we give the limiting magnitudes with various exposure time and signal-to-noise ratio for observers as references.
}
   \authorrunning{C.-H. Bai et al. }            
   \titlerunning{Upgraded Photometric System of 85-cm Telescope}  
   \maketitle
%
\section{INTRODUCTION}           
\label{sect:intro}
The 85-cm telescope is located at the Xinglong station (117$\degree$34$\arcminute$39$\arcsecond$ East, 40$\degree$23$\arcminute$26$\arcsecond$ North; at an altitude of about 960m) of National Astronomical Observatories of China (NAOC), jointly operated by Beijing Normal University and NAOC (\citealt{Fan:2016PASP}). The seeing values during 80\% of nights  are below 2.6\arcsecond with the distribution peak of around 1.8\arcsecond. The sky brightness at the zenith is around 21.1 mag arcsec$^{-2}$ in V-band. The average number of observable nights is 230 per year, and the average fraction of photometric nights is 32\% (\citealt{ZhangJC:2015}, \citealt{Zhang:2016PASP}). The main scientific projects include multi-color photometry of pulsating stars (\citealt{Luo:2012}), binaries (\citealt{YangYG:2013}, \citealt{ZhangYP:2015}, \citealt{ZhangLY:2015}), exoplanet (\citealt{liyunzhang:2015}), variable stars (\citealt{ZhangXB:2012}).

From 2014 to 2015, the telescope had been upgraded by the local technicians. It would be quite helpful and important if the observers could get to know the properties and performance of the telescope, including the bias, dark current and linearity of the CCD. In addition, analyzed results of the CCD photometric system are reported, such as the throughput, detection limit and instrument response.

We  introduce the 85-cm observation system in Section 2. Then the CCD characteristics are reported in Section 3. The photometric calibration and transformation coefficients are given in Section 4. The throughput and detection limit are addressed in Section 5, and a summary is given in Section 6.

\section{OBSERVATION SYSTEM}
\label{sect:Obs}
The 85-cm telescope is mounted equatorially. The effective diameter of the parabolic primary mirror is 850mm, concentrating 80\% of the energy into a circle with diameter of 1.6$\arcsecond$. The original focal length is 3000 $\pm$ 150mm. With the new corrector, the focal ratio of the prime focus reduces from 3.75 to 3.5. The new CCD camera is Andor iKon back-illuminated DZ936N - BV, 16 bit A/D converter, scientific-grade without anti-blooming gate, which was purchased by Department of Astronomy, Beijing Normal University and mounted to the telescope in Oct. of 2014. The specifications in the artificial manual are listed in Table 1. It has 5-stage peltier which can reach -80\textdegree C by air cooling or -95\textdegree C by coolant with recirculator. With a 2048 x 2048 imaging array (13.5 x 13.5 {$\mu$}m pixel$^{-1}$ ). The FOV of the CCD is 32$\arcminute$ x 32$\arcminute$ which is four times the size of the original 16.5$\arcminute$ x 16.5$\arcminute$ (\citealt{ZhouAY:2009}), and the pixel scale 0.93$\arcsecond$  is similar as the former reported. A Johnson-Cousins standard UBVIR filter system made by FLI  started to serve observations since Sep. of 2015. Time synchronization can be realized through the networks whenever needed.
\begin{center}
\begin{tabular}{lll}
\multicolumn{2}{c}{{\bf Table 1} Specifications of The CCD Camera According to The Product Manuals}\\
\hline
Features & Specifications \\
\hline
Pixel number    & 2048 x 2048  \\
Pixel size      & 13.5 {$\mu$}m x 13.5 {$\mu$}m  \\
Imageing area   & 27.6mm x 27.6mm   \\
Fill factor     & 100\%  \\
A/D conversion  & 16 bit  \\
Full well          & 100,000 e$^-$  \\
Scan rates                  & 50 kHz, 1 MHz, 3 MHz, 5 MHz (visualization mode)\\
Full frame readout time     & 91s@50kHz, 4.5s@1MHz, 1.65s@3MHz, 1.05s@5MHz\\
Operating temperature       & -80$\celsius$ Air cooled,$ $ -95$\celsius$ Coolant recirculator\\
                            & -100$\celsius$ Coolant chiller, coolant @ 10\celsius, 0.75 I/min\\
Readout noise               & 2.9e$^-$ @ 50 kHz, 7.0e$^-$ @ 1 MHz,\\
                            & 11.7e$^-$ @ 3 MHz, 31.5e$^-$ @ 5 MHz \\
Dark current e$^-$/pixel/sec    & 0.00040 @ -70\celsius, 0.00013 @ -80\celsius, 0.000059 @ -100\celsius \\
Linearity                       & $>$ 99 \% \\
\hline
\label{table:CCD}
\end{tabular}
\end{center}

\section{CCD CHARACTERISTICS}
\label{sect:CCD}
\subsection{Bias Level}
The bias level of the chip was tested at the CCD temperature of -70\celsius. We took the bias images during the night of 11th Oct. 2017. During the test, environment temperature changed from 3{\celsius} to 6{\celsius} as shown in Figure 1. One can see that the bias is very stable with the mean value of about 300 $\pm$ 2 Analog-to-Digital (ADU).  The scan rates of the chip was set to 1 MHz with the gain of 1x. Table 2 lists the test results.

\begin{figure}
  \begin{center}
        \scalebox{0.5}[0.5]{\includegraphics*[angle=-90]{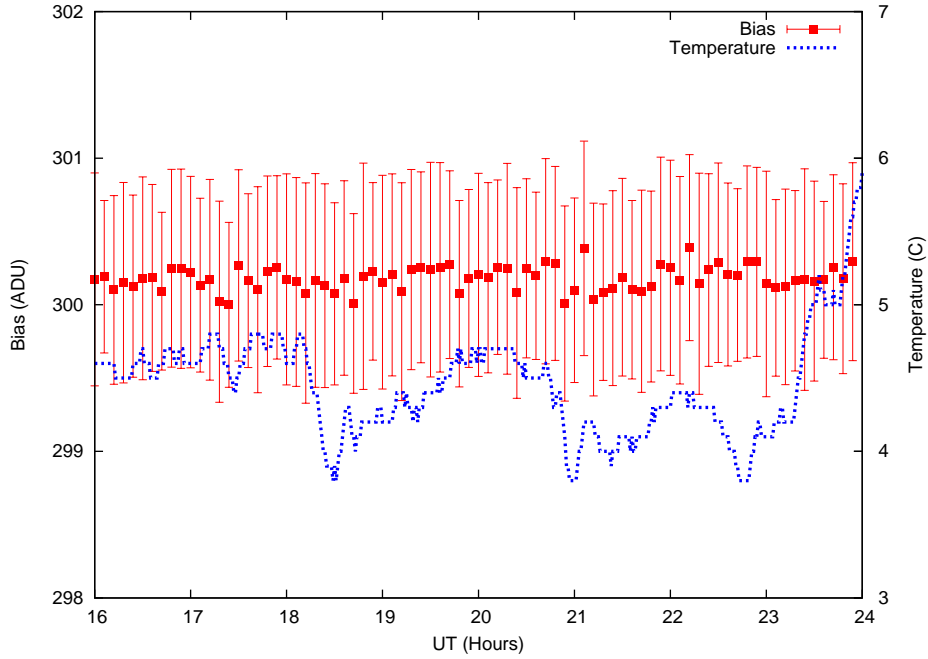}}
        \caption[Fig. Bias]{{\small Bias level at the temperature of -70\celsius, with the scan rates of 1 MHz, Gain of 4x. The blue curve shows the temperature variations.}}
  \end{center}
    \label{fig:Bias_temp}  
\end{figure}
\begin{center}
\begin{tabular}{lll}
\multicolumn{2}{c}{{\bf Table 2} Test Results About Bias, Gain, Readout Noise, Dark and Linearity}\\
\hline
Bias (ADU)                  & mean value 300 $\pm$ 3 @ 1 MHz, Gain 1x\\
Gain                        & 0.97 @ 1x, 1.88 @ 2x, 3.43 @ 4x, @ 1 MHz\\
Readout noise               & 6.8e$^-$ @ 1 MHz, Gain 1x \\
Dark current e$^-$/pixel/sec    & 0.00448 $\pm$ 0.00027 @ -70\celsius \\
linearity                       & $>$ 99.6 \% \\
\hline
\label{table:test}
\end{tabular}
\end{center}

\subsection{Gain and Readout Noise}
The gain G is the factor that converts the digital output to the number of electrons , which is useful to evaluate the performance of the CCD camera. We measured the gain by comparing the signal level to the amount of variation in the flat field and bias images. The gain and readout noise can  be measured by equations \ref{equ:gain} and \ref{equ:RN} (\citealt{Howell:2000}, \citealt{Howell:2006})  with IRAF's (Image Reduction and Analysis Facility, provided and maintained by NOAO) task of {\it findgain}. Task {\it findgain} uses  these equations to calculate the gain and readout noise with two dome flats and two bias frames. When we select CCD gain 4x and scan rates 1 MHZ which are taken by most observers, the  test value of gain and readout noise are 0.97 e$^-$/{\it ADU} and 6.8 e$^-$ respectively. The results approach the values provided by the artificial manual of 1 e$^-$/{\it ADU} and 7.0 e$^-$, respectively, as listed in Table 1.
\begin{eqnarray}
Gain = \frac{(\overline{F1}+\overline{F2})-(\overline{B1}+\overline{B2})}{\sigma_{F1-F2}^{2} - \sigma_{B1-B2}^{2}}
\label{equ:gain}
\end{eqnarray}

\begin{eqnarray}
Read out noise = \frac{ Gain \cdot \sigma_{\overline{B1} - \overline{B2}}}{\sqrt {2}}
\label{equ:RN}
\end{eqnarray}

\subsection{Dark current}
\begin{figure}
    \begin{minipage}[t]{0.5\linewidth}
        \scalebox{0.3}[0.3]{\includegraphics*[angle=-90]{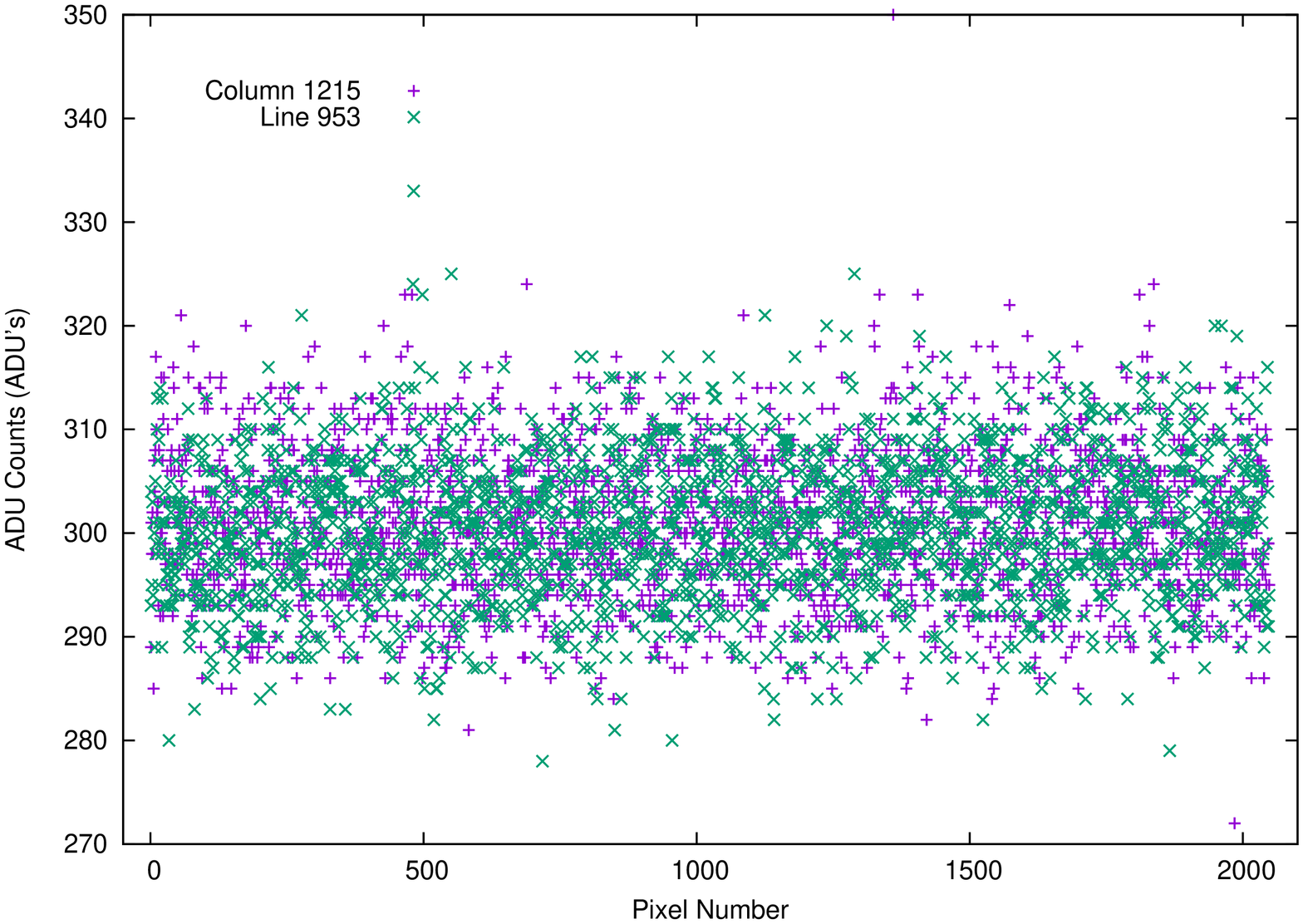}}
        \caption[Fig. Dark pix]{{\scriptsize ADU distribution of a 1200s dark image at the CCD temperature of -70\celsius, with the scan rates of 1 MHz in arbitrary column and line.}}
    \label{fig:Dark_pixel}
    \end{minipage}
    \begin{minipage}[t]{0.5\linewidth}
        \scalebox{0.3}[0.3]{\includegraphics*[angle=-90]{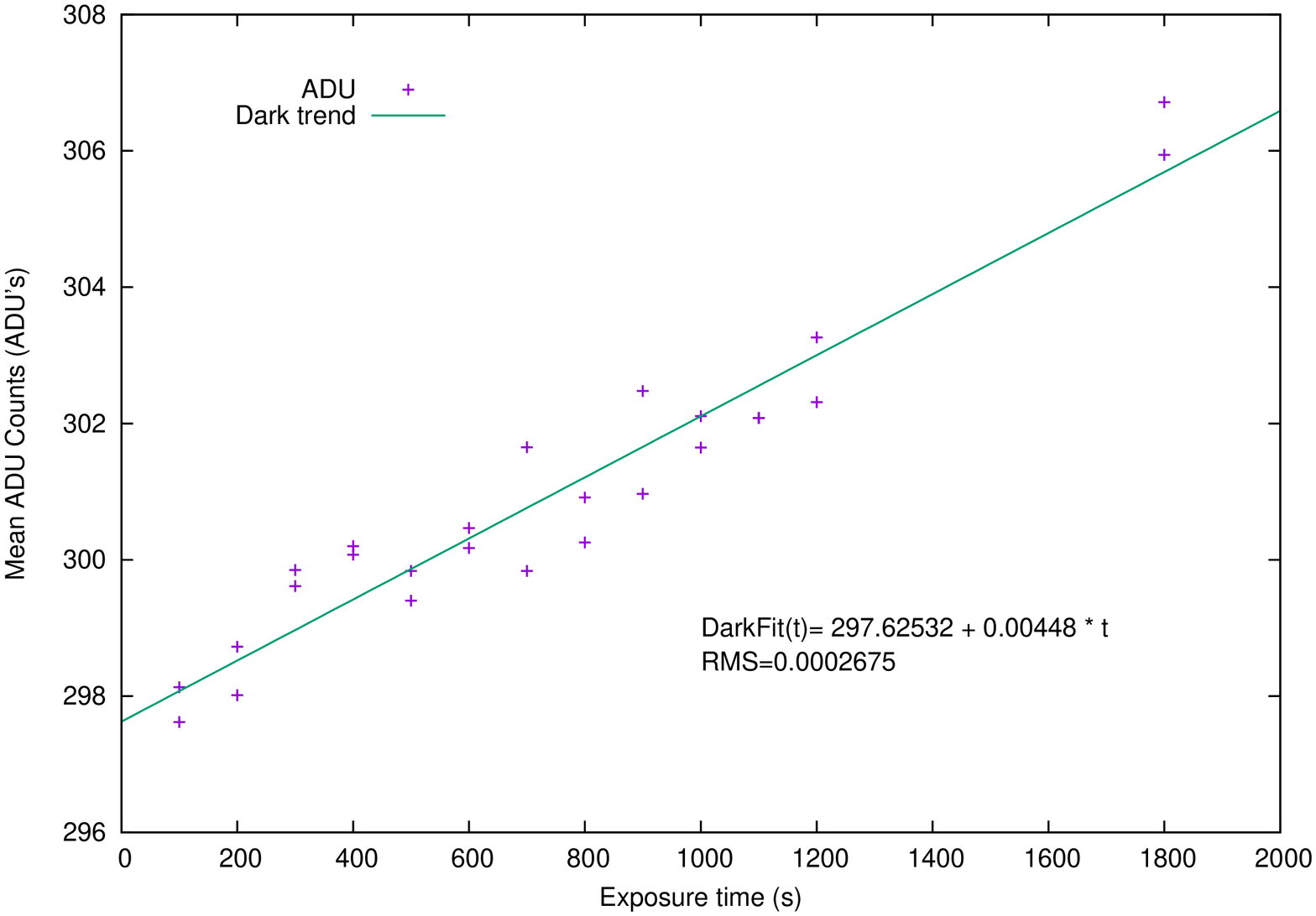}}
        \caption[Fig. Dark trend]{{\scriptsize The average ADU count versus exposure time of dark images. The solid line is the best linear fit
        between the average ADU count and the exposure time.}}
    \end{minipage}
    \label{fig:Dark_trend}
\end{figure}

The dark current measurement was taken for one whole night on Apr. 14th 2016. The dark counts with 1200s of exposure time were stable in columns and lines as show in Figure \ref{fig:Dark_pixel}. The dark images were taken with the exposure time from 100s to 1800s. For each exposure time 2 frames were taken. Hence, we got the mean and standard deviations from each image. As shown in Figure 3, the measured dark current is 0.00448 $\pm$ 0.00027 e$^-$/pixel/sec. This value seems to be much larger than that given in the product manual (0.00040 e$^-$/pixel/sec @ -70\celsius), but the impact on the practical observation is very small. Therefore dark correction is unnecessary for  short  exposure which is thorter than 600s.

\subsection{Linearity}
The linearity of CCD is  important for high precision photometry. In order to check the linearity of the response of the CCD, we carried out measurements with dome flat in two nights of May 2016. As the stray light  changes with the sun during daytime, the measurements were taken after the dusk. We closed the shutter and adjusted the dome light by covering with several pieces of A4 paper, which made it dim enough for the CCD camera to take long time exposure.

At the beginning, we set different CCD exposure time to make the ADU values larger than 60000,  far from saturated with Gain 4x and scan rate 1MHz. The ADU values became smaller when we changed the gain set to 2x and 1x. We could get the real values through the images, which were corrected with bias, as listed in Table 3.
\begin{center}
\begin{tabular}{cccc}
\multicolumn{4}{c}{{\bf Table 3} Gain values with scan rate @ 1 MHz  }\\
\hline
Exposure time & Gain selected & mean ADU & Gain value\\
\hline
30 s & 1 x & 18284 & 3.43\\
30 s & 2 x & 33468 & 1.88\\
30 s & 4 x & 62863 & 0.97\\
\hline
\label{table:linearity}
\end{tabular}
\end{center}

Then, we pointed  telescope  near the zenith and carried out two groups of linearity measurement with the dome light. For each exposure time, we took 6 frames. The mean ADU values of a 400 x 400 region close to  the CCD center were counted. For one group with the CCD scan rate of 1MHz and the gain of 1x, we set the exposures from 0.1s to 70s. The mean count increased with the exposure time, but the standard deviation droped after CCD reached the full well. The linearity peak of gain 1x is larger than 29000, as shown in Figure 4, which multiplied with the gain value 3.43 equals to 99470 ( $>$ 99000 ). One can find that the linearity is better than 99\% for the full well. For another group with scan rate 1MHz and gain 4x, the exposure time was set from 0.5s to 46s. When the gain is 4x (real value is 0.97), the 16 bit ADU value falls down after the AD converter saturated at 65535 as shown in Figure 5. We understand that the electron count is far from the full well of 100000. At the low edges of the data points in the two figures, the mean value deviated from the fit line, which is due to the shutter effect. Therefore, we suggest the exposure time larger than 3s.

\begin{figure}
    \begin{minipage}[t]{0.5\linewidth}
        \scalebox{0.25}[0.25]{\includegraphics*{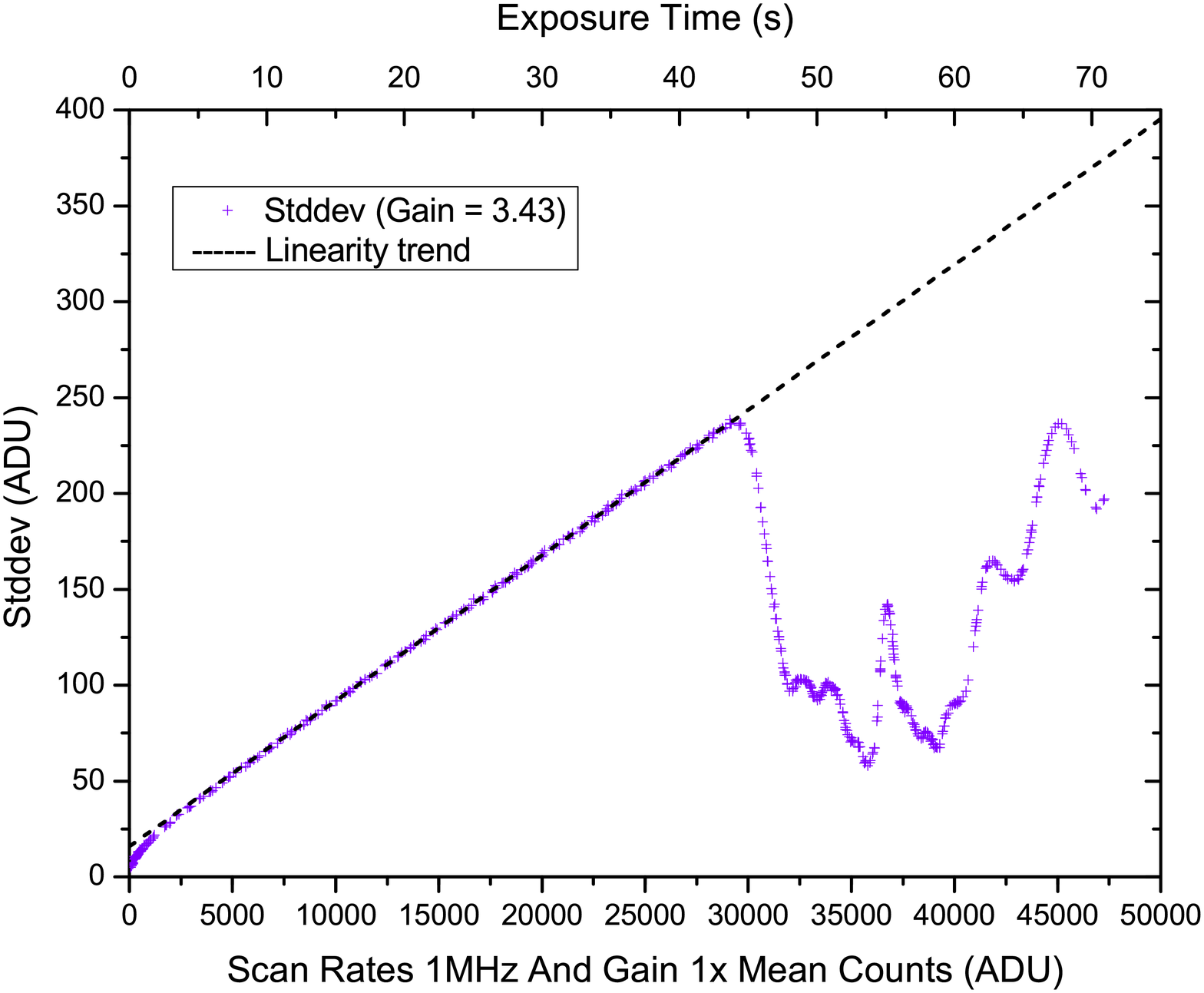}}
        \caption[Fig. linearity3.4]{{\scriptsize Plot of the standard deviation versus the mean count of the dome flat with CCD scan rate of 1MHz and gain 1x. }}
    \label{fig:linearityG1}   
    \end{minipage}
    \begin{minipage}[t]{0.5\linewidth}
        \scalebox{0.3}[0.3]{\includegraphics*{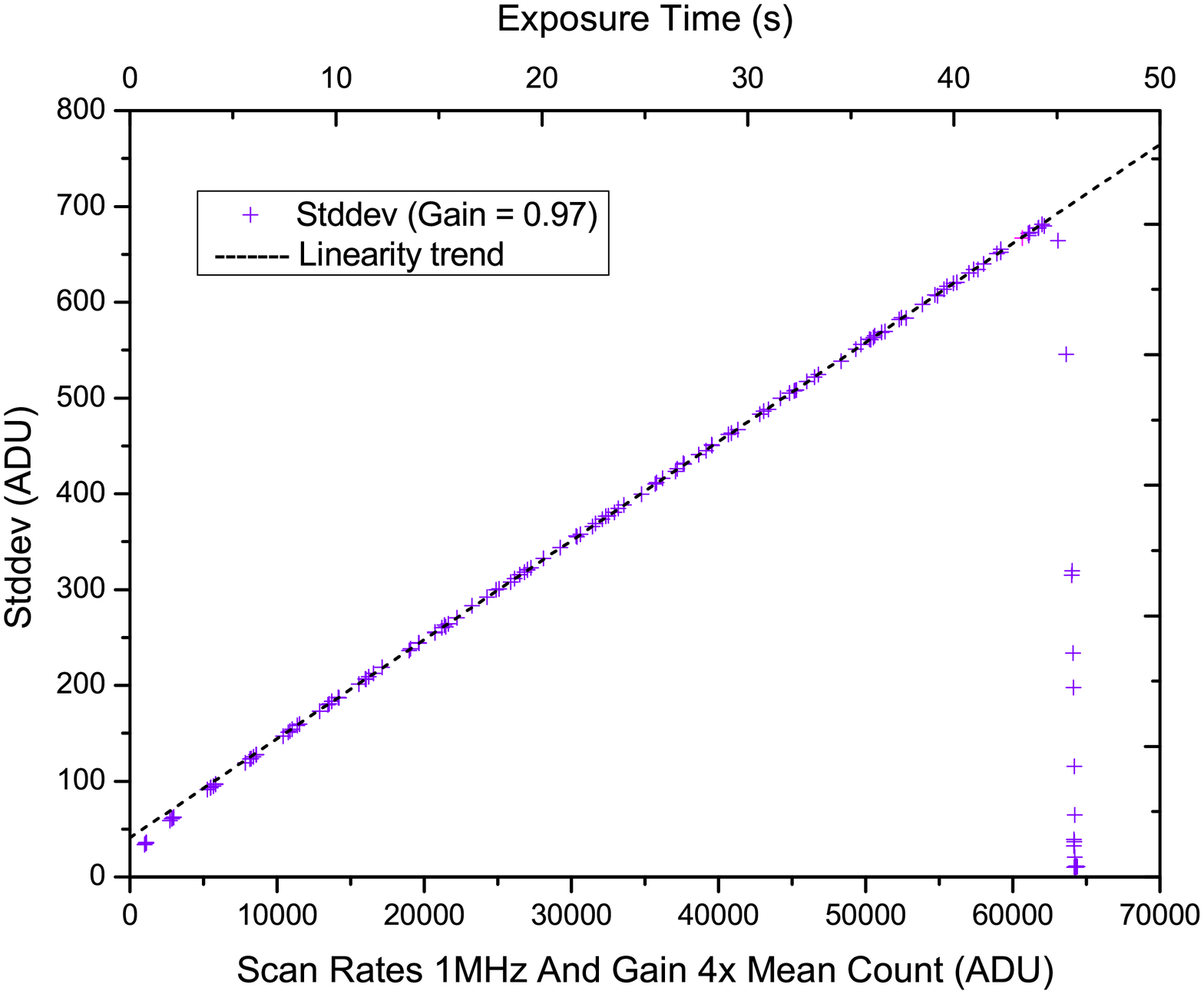}}
        \caption[Fig. linearity0.97]{{\scriptsize Plot of the   standard deviation versus the mean count of the dome flat with CCD scan rate of 1MHz and gain 4x. }}
    \end{minipage}
    \label{fig:linearityG4}
\end{figure}

\section{PHOTOMETRIC CALIBRATION}
\label{sect:PHOTOMETRIC}
The magnitudes measured from the images are the telescope instrumental magnitudes. In order to compare the photometric result with that from the other instruments, one needs to calibrate the magnitudes from the instrumental system into the standard system. In Feb. Mar. and Apr. 2016, we selected some stars from Landolt standards (\citealt{landolt:1992}, \citealt{landolt:2013}) as listed in Table 4, considering both the magnitude and colors. Some of them  can pass the zenith at Xinglong station, hence are suitable for calibration.
\begin{center}
\begin{tabular}{lcccccccc}
\multicolumn{9}{c}{{\bf Table 4} Landolt Standards Used in 2016}\\
\hline
Star-Name & $\alpha$ (2000) & $\sigma$ (2000) & \it{V} & \it{B - V} & \it{U - B} & \it{V - R} & \it{R - I} & \it{V - I}\\
\hline
    104 456 & 12:42:53 & -00:32:09 & 12.362 &  0.622 &  0.135 &  0.357 &  0.337 &  0.694 \\
     94 394 & 02:56:14 & +00:35:25 & 12.273 &  0.545 & -0.047 &  0.344 &  0.330 &  0.676 \\
     95 328 & 03:54:19 & +00:36:47 & 13.525 &  1.532 &  1.298 &  0.908 &  0.868 &  1.776 \\
      97 75 & 05:57:55 & -00:09:07 & 11.483 &  1.872 &  2.100 &  1.047 &  0.952 &  1.999 \\
     G44 27 & 10:36:02 & +05:07:11 & 12.636 &  1.586 &  1.088 &  1.185 &  1.526 &  2.714 \\
PG1047+3C   & 10:50:18 & -00:00:21 & 12.453 &  0.607 & -0.019 &  0.378 &  0.358 &  0.737 \\
SA 29-322   & 09:46:31 & +44:22:32 &  9.766 & +0.488 & +0.030 & +0.285 & +0.262 & +0.560 \\
SA 32-377   & 12:55:45 & +44:40:38 & 10.630 & +0.641 & +0.101 & +0.372 & +0.358 & +0.735 \\
SA 35-343   & 15:51:57 & +44:35:26 & 10.870 & +0.433 & -0.063 & +0.271 & +0.254 & +0.529 \\
SA 38-358   & 18:48:23 & +45:23:22 &  9.854 & +1.132 & +1.186 & +0.583 & +0.509 & +1.095 \\
SA 29-324   & 09:46:53 & +44:25:05 & 11.304 & +1.117 & +1.075 & +0.582 & +0.516 & +1.097 \\
SA 32-330   & 12:55:26 & +44:33:35 & 10.068 & +0.665 & +0.190 & +0.378 & +0.342 & +0.721 \\
SA 35-339   & 15:51:54 & +44:32:29 & 12.775 & +0.554 & +0.028 & +0.336 & +0.329 & +0.662 \\
SA 38-365   & 18:48:32 & +45:25:10 & 11.625 & +1.205 & +1.309 & +0.616 & +0.546 & +1.160 \\
\hline
\label{table:obsed}
\end{tabular}
\end{center}

The seeing values approximated 2.2\arcsecond during the observation night. After colleting the  photometric data for the standars stars, all frames were reduced with 2 times aperture by standard photometry steps using the {\it apphot} package of IRAF. We take 25 as our instrumental magnitudes zero point, then the instrumental magnitudes and airmass are obtained. We define transformation equations offered by IRAF as follows,
\begin{eqnarray}
 U_{inst}&=&U_{std}+Z_U+K^{'}_{U}X+C_U(U-B)_{std} \\
 B_{inst}&=&B_{std}+Z_B+K^{'}_{B}X+C_B(B-V)_{std} \\
 V_{inst}&=&V_{std}+Z_V+K^{'}_{V}X+C_V(B-V)_{std} \\
 R_{inst}&=&R_{std}+Z_R+K^{'}_{R}X+C_R(V-R)_{std} \\
 I_{inst}&=&I_{std}+Z_I+K^{'}_{I}X+C_I(V-I)_{std}
\end{eqnarray}
where U$_{std}$, B$_{std}$, V$_{std}$, R$_{std}$ and I$_{std}$ are the standard magnitudes, U$_{inst}$, B$_{inst}$, V$_{inst}$, R$_{inst}$, I$_{inst}$, are the instrumental magnitudes, Z$_U$, Z$_B$, Z$_V$, Z$_R$ and Z$_I$ are zero point magnitudes, K$^{'}_{U}$, K$^{'}_{B}$, K$^{'}_{V}$, K$^{'}_{R}$ and K$^{'}_{I}$ are the first-order extinction coefficients, C$_U$,  C$_B$, C$_V$ ,C$_R$ and C$_I$ are the color terms in the transformation equations , and X denotes the airmass.

For each band we have obtained 40 frames. Then, we use the IRAF's task of {\it mknobsfile} and {\it fitparams}  in {\it photcal} package to fit the observed magnitudes and derive the coefficients. The extinction coefficients (K$^{'}$) and color terms (C) are fitted separately. The second-order extinction terms are too small to be ignored. The obtained transformation coefficients  are listed in Table 5, including the filter names, zero points (Z), first-order extinction coefficients (K$^{'}$), color terms (C) and RMS. In Table 6, we  compare our results with historical measurements at Xinglong station.

\begin{center}
\begin{tabular}{ccccc}
\multicolumn{5}{c}{{\bf Table 5} Coefficients, Standard deviation and RMS of 85-cm telescope photometric system}\\
\hline
Filter & Zero points (Z) & Extinction ($K^{'}$) & Color Term (C) & RMS\\
\hline 
U$_{inst}$ & 4.732 $\pm$ 0.032 & 0.590 $\pm$  0.022 & -0.376 $\pm$ 0.034 & 0.051 \\
B$_{inst}$ & 1.781 $\pm$ 0.042 & 0.431 $\pm$  0.029 & -0.108 $\pm$ 0.011 & 0.043 \\
V$_{inst}$ & 1.909 $\pm$ 0.037 & 0.282 $\pm$  0.026 & -0.088 $\pm$ 0.025 & 0.052 \\
R$_{inst}$ & 1.977 $\pm$ 0.027 & 0.217 $\pm$  0.019 & -0.145 $\pm$ 0.008 & 0.049 \\
I$_{inst}$ & 2.383 $\pm$ 0.029 & 0.156 $\pm$  0.021 & -0.077 $\pm$ 0.008 & 0.050 \\
\hline
\label{table:coef}
\end{tabular}
\end{center}

\begin{center}
\begin{tabular}{ccccccc}
\multicolumn{7}{c}{{\bf Table 6} Extinction Coefficiats at Xinglong station }\\
\hline
Year        & K$^{'}_{U}$        & K$^{'}_{B}$        & K$^{'}_{V}$        & K$^{'}_{R}$        & K$^{'}_{I}$        & Ref\\
\hline
2016        & 0.590 $\pm$  0.022 & 0.431 $\pm$  0.029 & 0.282 $\pm$  0.026 & 0.217 $\pm$  0.019 & 0.156 $\pm$  0.021 & this paper\\
2011-2012   &  & 0.348$\pm$0.022 & 0.236$\pm$0.017 & 0.168$\pm$0.019 & 0.085$\pm$0.021 &  (\citealt{Huang:2012RAA})\\
2008        &  & 0.330$\pm$0.007 & 0.242$\pm$0.005 & 0.195$\pm$0.004 & 0.066$\pm$0.003 &  (\citealt{ZhouAY:2009})\\
2006-2007   &  & 0.307$\pm$0.009 & 0.214$\pm$0.008 & 0.161$\pm$0.008 & 0.091$\pm$0.008 &  (\citealt{Huang:2012RAA})\\
2004-2005   &  & 0.296$\pm$0.012 & 0.199$\pm$0.009 & 0.141$\pm$0.010 & 0.083$\pm$0.009 &  (\citealt{Huang:2012RAA})\\
1995        &  & 0.35 & 0.20 & 0.18 & 0.16 &  (\citealt{Shi:1998ChAA})\\
1998        &  & 0.31 & 0.22 & 0.14 & 0.10 &  (\citealt{Shi:1998ChAA})\\
\hline
\label{table:coef_his}
\end{tabular}
\end{center}

 The comparison between the Landolt standard magnitudes and the calibrated magnitudes derived from the transformation equations in UBVRI bandsis shown in Figure 6.
\begin{figure}
    \begin{center}
    \scalebox{0.4}[0.4]{\includegraphics*[angle=-90]{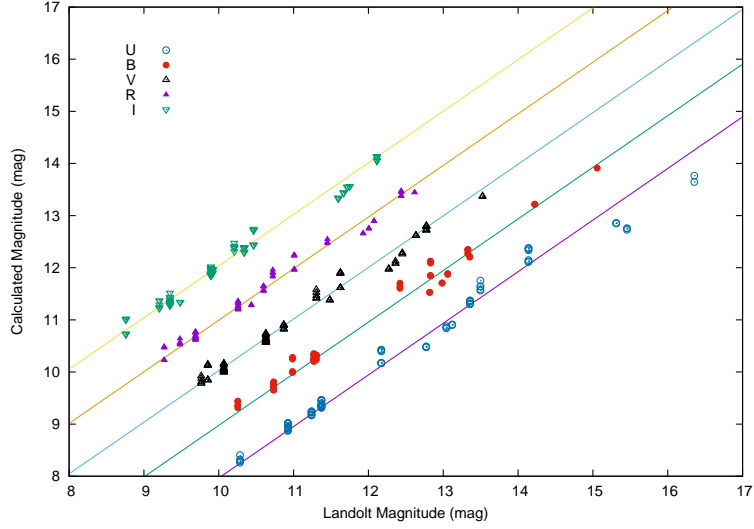}}
    \caption[Fig. 4.1]{{\scriptsize Landolt magnitudes versus calculated magnitudes using derived transformation coefficients in {\it UBVRI} bands. The U, B, R, I calculated magnitudes are shifted to -2, -1, +1, +2
    magnitudes, Respectively, for clarify. Solid lines are the best linear fits.}}
    \end{center}
    \label{fig:stand_coeff}  
\end{figure}

With the relationship between the Landolt standard magnitudes and the shifted calibrated magnitudes using the transformation coefficients derived from the observations taken on Feb. Mar. and Apr. 2016, the slopes of all the fitted lines are close to 1.0, so the calibration is good. The magnitude rms in each filter is smaller than 0.1 as listed in Table 6, indicatting that the calibration relationship between the photometric system of 85-cm and the Johnson-Cousins standard photometric system is well established.

\section{SYSTEM PERFORMANCE}
\label{sect:PERFORMANCE}

\subsection{System Throughput}
The total throughput of entire optical system can be estimated by observing a number of Landolt standards. The throughput includes the primary mirror reflection, correctors, filter response, quantum efficiency of the detector and the atmospheric transformation. Kinoshita has introduced his method (\citealt{Kinoshita:2005ChJAA}) worked at Lulin observatory. We calculate the throughput efficiency of the 85-cm telescope by another way which was introduced by Fan et al. (\citealt{Fan:2016PASP}) based on the flux of Vega (\citealt{bohlin:2004}, \citealt{bohlin:2014}) following Equation (8).
\begin{eqnarray}
\eta(\lambda) = \frac{F_{ADU}\cdot G}{F_{\lambda} \cdot \delta\lambda \cdot S_{tel}}
\end{eqnarray}

Where F$_{ADU}$ is the observed number counts of a standard star per second (ADU s$^{-1}$); G  the gain of the CCD (e$^{-1}$ ADU$^{-1}$); F$_{\lambda}$  theoretical photon flux of a standard star derived from its AB mag ( photon s$^{-1}$ cm$^{-2}$ {\AA}$^{-1}$); $\delta\lambda$
 the effective band width in the filter in imaging observations or the dispersion of the grating for spectroscopic observations ({\AA}); S$_{tel}$  the effective area of the primary mirror of the telescope (cm$^{2}$) and $\lambda$ is the effective wavelength of the filter or the wavelength at which the efficiency is  computed for the spectroscopy ({\AA}).

The median throughput in different bands are listed in Table 7.
\begin{center}
\begin{tabular}{cccccc}
\multicolumn{6}{c}{{\bf Table 7} The total throughput of 85-cm in UBVRI bands}\\
\hline
Band & U & B & V & R & I \\
\hline
Throughput & 4.5\% & 23.5\% & 33.8\% & 28.2\% & 27.8\%\\
\hline
\label{table:througt}
\end{tabular}
\end{center}

\subsection{Limiting Magnitude And Photometry Accuracy}
The sky brightness has been introduced in detail by Zhang et al. (2015), which is  deeper than the 85-cm limiting detection. We can use Equation (9) to make an estimation(\citealt{Howell:2000}).
\begin{eqnarray}
SNR &=& \frac{N_{star}}{\sqrt{N_{star}+n_{pix}(N_{sky}+N_{dark}+N_{readout}^{2})}} \\
\sigma &=& 1.0857 / SNR .
\end{eqnarray}

 Where N$_{star}$ is the total number of target photons collected by the CCD camera, n$_{pix}$  the number of pixels which  change with the seeing condition. N$_{sky}$  the total number of photons on each pixel from the sky background, N$_{dark}$  the dark current per pixel, N$_{readout}$  the readout noise estimated in Section 3.2. The dark current effect is neglected here. We can easily get the photometric error from the SNR by the equation above, where $\sigma$  in Equation (10) is the photometric error for objects. 1.0857 is the correction term between an error in flux electron and  same error in magnitude.

\begin{figure}
    \begin{minipage}[t]{0.5\linewidth}
    \scalebox{0.3}[0.3]{\includegraphics*[angle=-90]{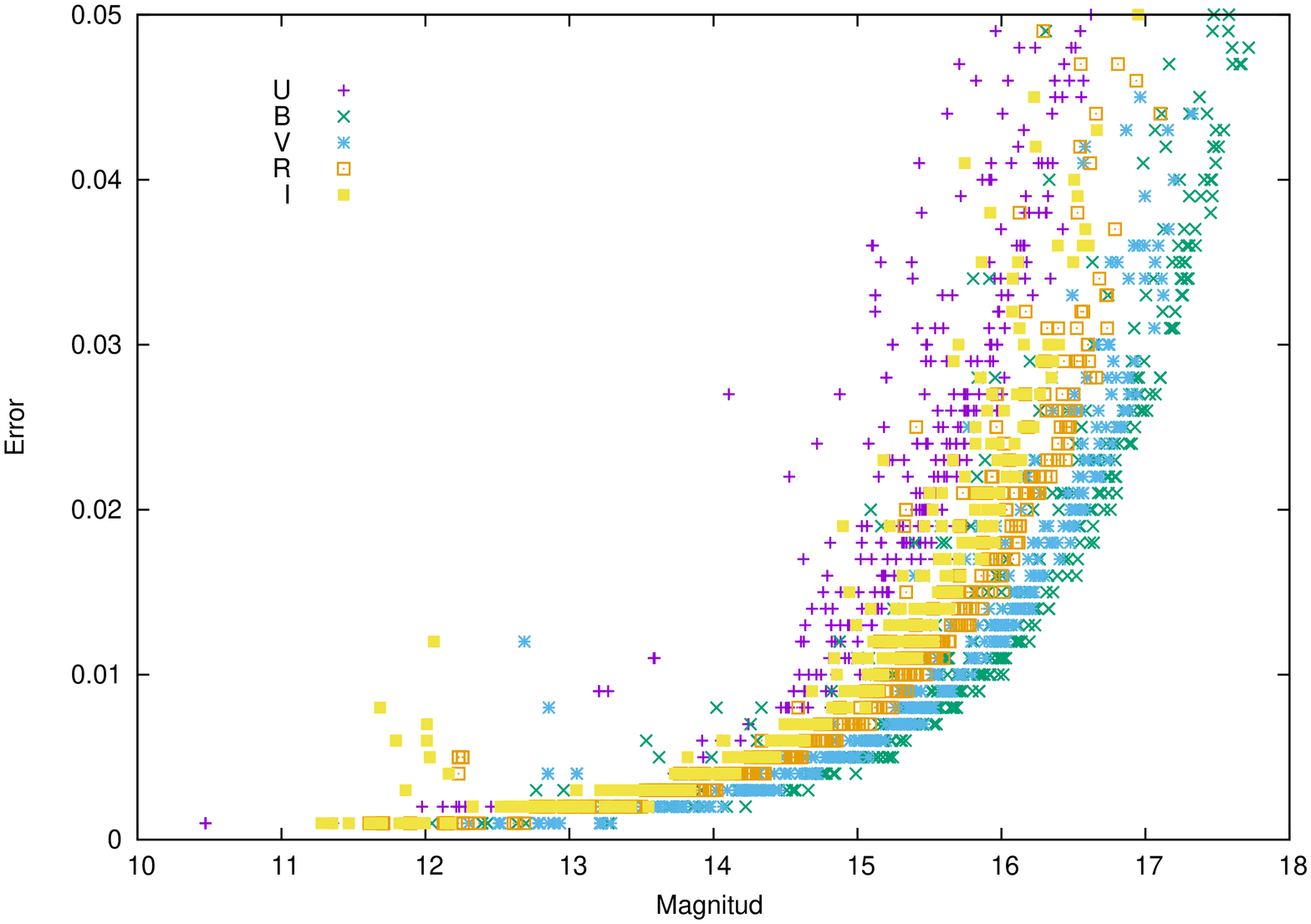}}
    \caption[Fig. m92]{{\scriptsize Photometry error versus calibrated magnitude of M92 with 600 s for U, 160 s for B, 144 s for V, 72 s for R and I band, respectively.}}
    \label{fig:ubvrim92}  
    \end{minipage}
    \begin{minipage}[t]{0.5\linewidth}
    \scalebox{0.3}[0.3]{\includegraphics*[angle=-90]{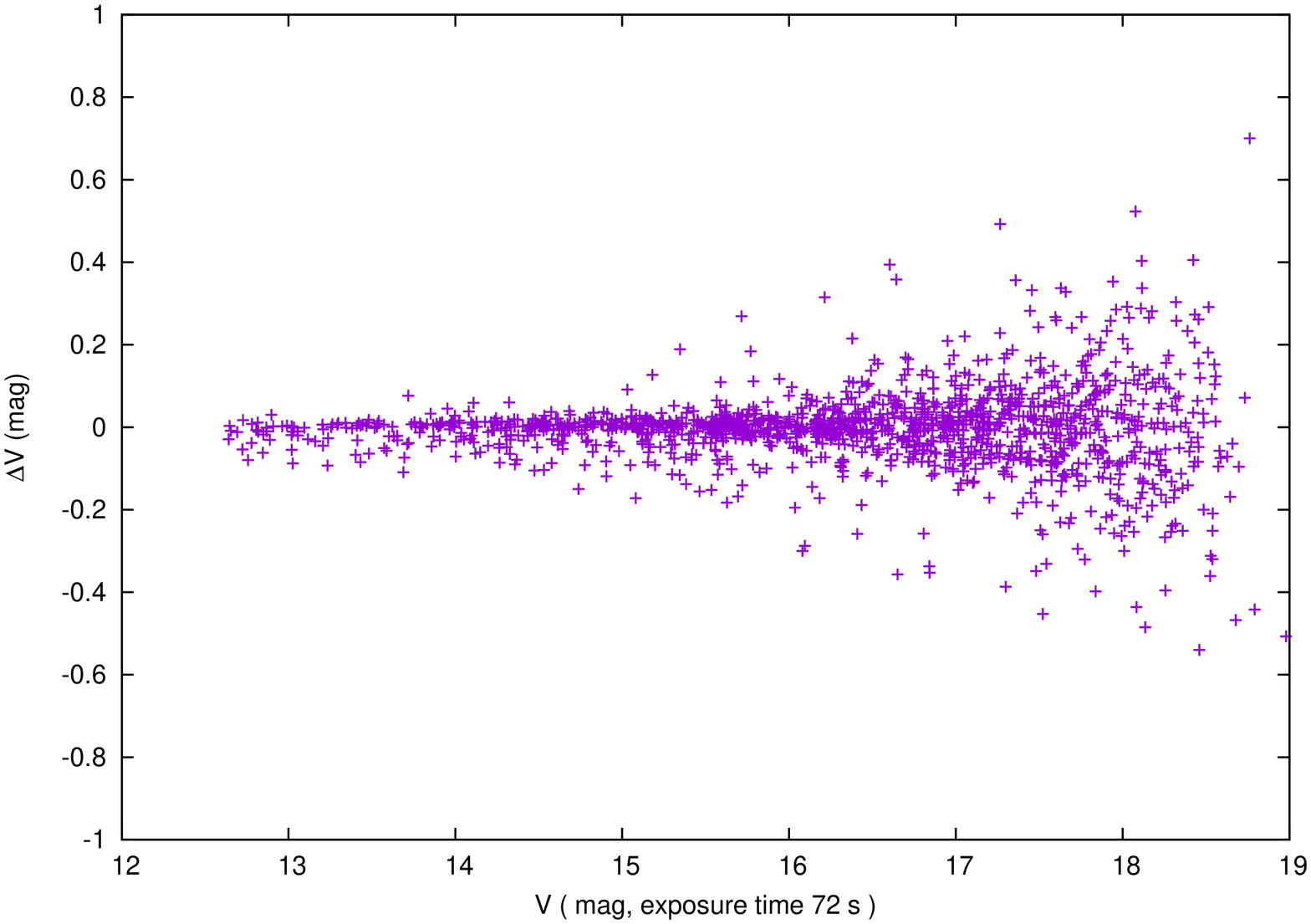}}
    \caption[Fig. accuracy]{{\scriptsize Photometry accuracy in V  band with the exposre time of 72 s.}}
        \end{minipage}
    \label{fig:accu}  
\end{figure}

To check the reliability of the calibration relationship, we carried out  observations for the globular cluster M92 for one night in June 2016. Here, the relations among the exposure time, the SNR and the star magnitude which we obtained are based on IRAF photometric results. To avoid saturation and get higher SNR, the exposure time was 600 s for U, 160 s for B, 144 s for V, 72 s for either R or I band. The relationship between the brightness and the photometry error is shown in Figure \ref{fig:ubvrim92}. We took two 72 s exposures for M92, then photometry of all stars with SNR higher than 2 was made. The brightness versus photometry accuracy $\delta$ magnitude of the two measurements was plotted in Figure 8, it is similar to the result of Weihai Observatory(\citealt{Hu:2014RAA}).

During the test night, we made 5 sets of different exposure time in each filter. The airmass was changing during the observations. After the standard reduction, we get the time, magnitude, photometry  accuracy with different arimass in each band. So the relationship between the signal-to-noise ratios of the limiting magnitudes which changed with exposure time and the airmass can be derived. In Figure 9 and 10, we plot the limiting magnitude with SNR at 100 and 200  versus the exposure time. Observers can use these figures to estimate the exposure time of their targets. Based on our observations, a simulation was done with a 300s exposure at the  zenith. The limiting magnitudes with the SNR of 5 and the exposure of 300s are 18.7, 20.9, 21.1, 20.6, 20.3 mag in the U, B, V, R and I bands, respectively.
\begin{figure}
    \begin{minipage}[t]{0.5\linewidth}
        \scalebox{0.3}[0.3]{\includegraphics*[angle=-90]{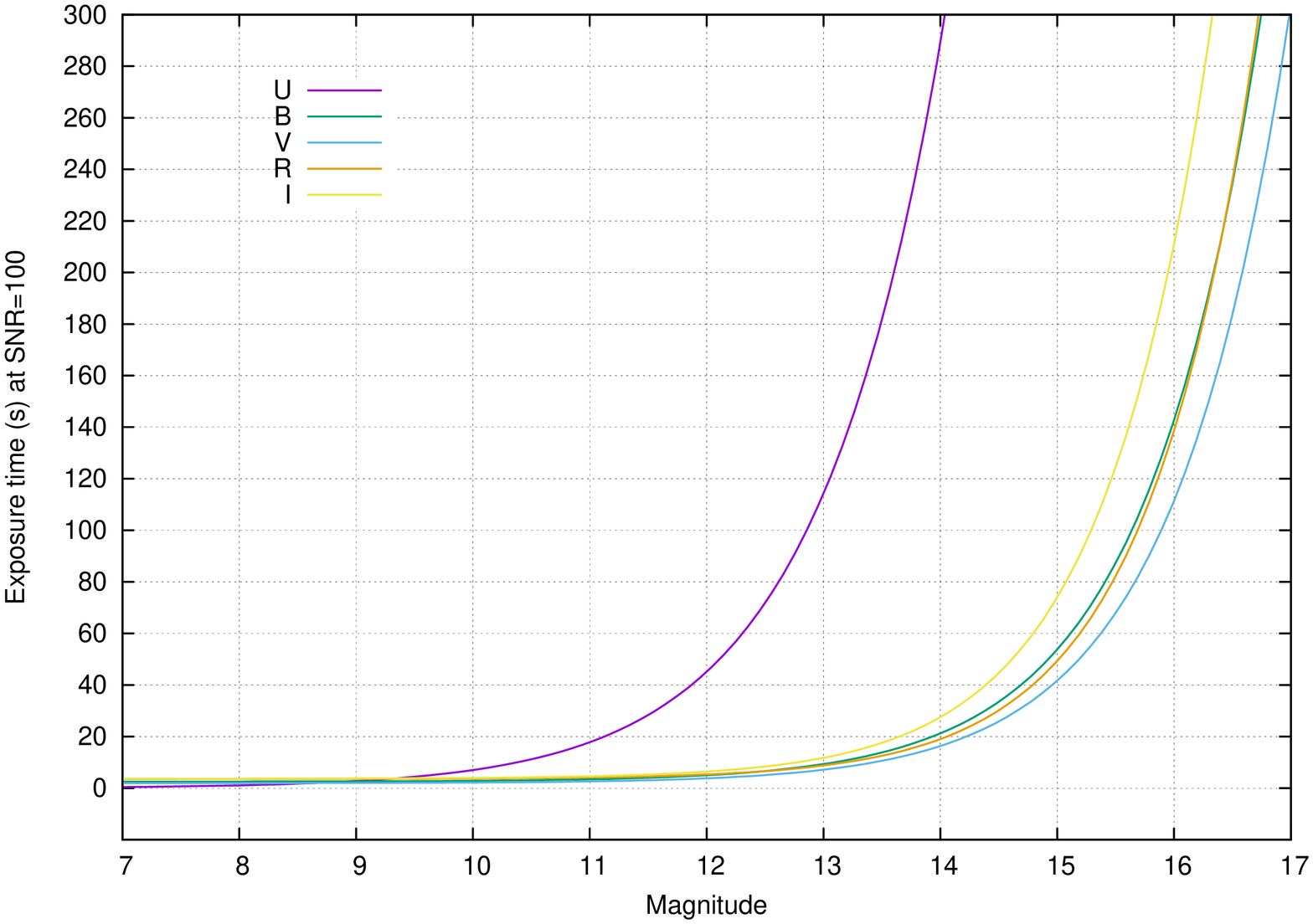}}
        \caption[Fig. SN100]{{\scriptsize Exposure time with limiting magnitude at SNR = 100.}}
    \label{fig:SN100}
    \end{minipage}
    \begin{minipage}[t]{0.5\linewidth}
        \scalebox{0.3}[0.3]{\includegraphics*[angle=-90]{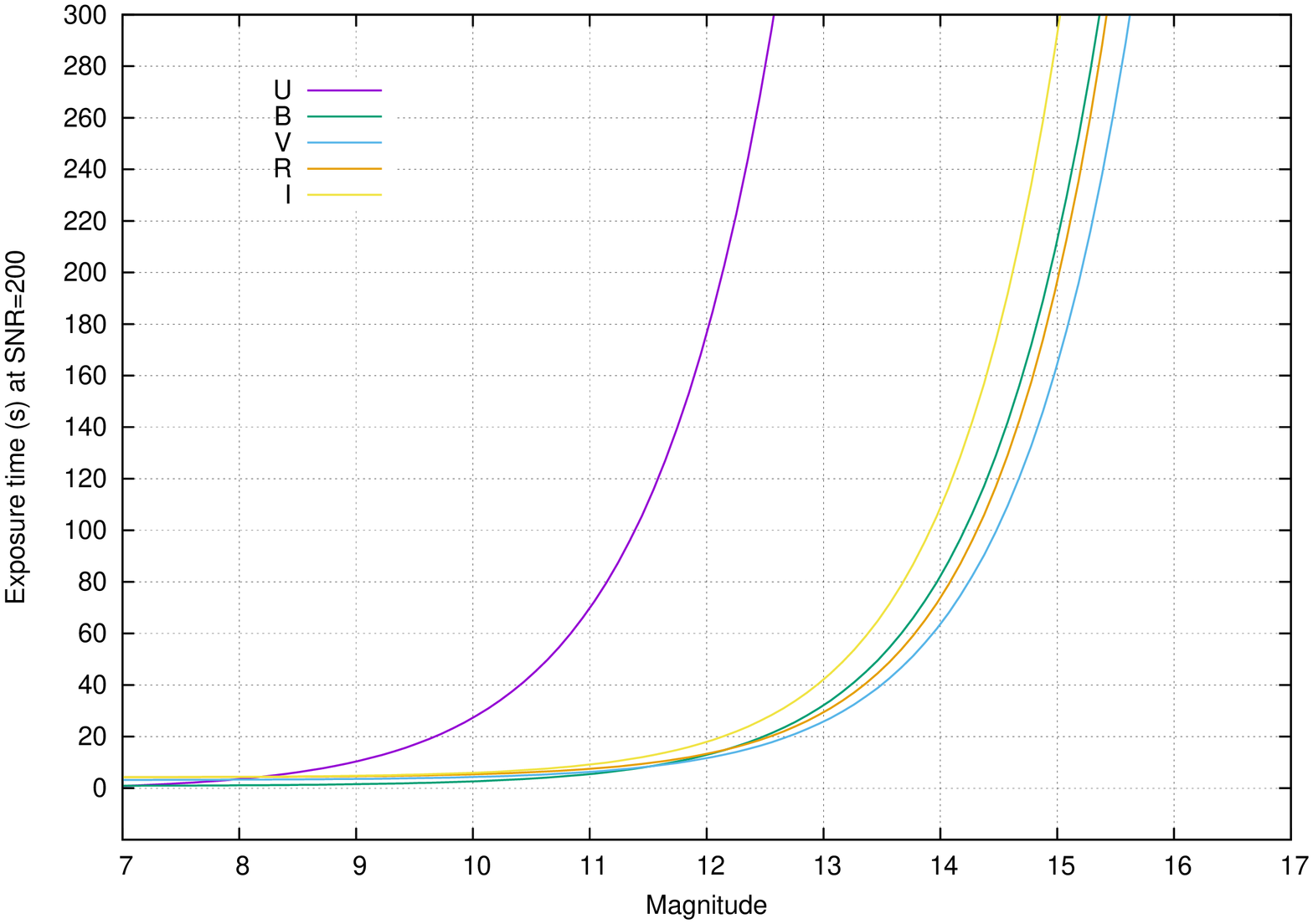}}
        \caption[Fig. SN100]{{\scriptsize Exposure time with limiting magnitude at SNR = 200.}}
    \end{minipage}
    \label{fig:SN200}
\end{figure}

\section{SUMMARY}
\label{sect:SUMMARY}
In this paper, we introduce the photometric system of the upgraded 85-cm telescope. In order to offer more valuable information to observers, we test the bias, gain, readout noise, dark current and linearity of the CCD camera which is mounted at the prime focus. The results show that it is a very good
scientific-grade camera. Based on several nights of observations for several Landolt standard, the transformation coefficients are derived between the instrumental UBVRI magnitude and the standard magnitude. The atmospheric extinction coefficients and color terms are found to be similar as the historical measurements.
Due to the new corrector and the CCD camera, the throughput of the 85-cm telescope is better than the other small size telescopes located at Xinglong station.
The limiting magnitude can reach 18.7, 20.9, 21.1, 20.6, 20.3 mag in U, B, V, R and I band with the SNR of 5 and the exposure of 300s, respectively. At last, we calculate the limiting magnitude versus the exposure time at different SNR  to give some exposure advices.

\begin{acknowledgements}
We gratefully acknowledge the support of the staff of the Xinglong 85-cm telescope. JNF acknowledges the support from the National Natural Science Foundation of China (NSFC) through the grant 11673003 and the National Basic Research Program of China (973 Program 2014CB845700). This work was partially supported by the Open Project Program of the Key
Laboratory of Optical Astronomy(National Astronomical Observatories), National Natural Science Foundation of China No. 11403088 and 11273051 (Xinjiang Astronomical Observatory), Chinese Academy of Sciences and Department of Astronomy, Beijing Normal University.
\end{acknowledgements}

\bibliographystyle{raa}
\bibliography{85cm_raa_R3_clean_sbm}

\begin{thebibliography}{20}
\providecommand\natexlab[1]{#1}
\providecommand\JournalTitle[1]{#1}

\bibitem[{Bohlin}(2014)]{bohlin:2014}
{Bohlin}, R.~C. 2014, AJ, 147, 127

\bibitem[{Bohlin} \& {Gilliland}(2004)]{bohlin:2004}
{Bohlin}, R.~C., \& {Gilliland}, R.~L. 2004, AJ, 127, 3508

\bibitem[{Fan} {et~al.}(2016)]{Fan:2016PASP}
{Fan}, Z., {Wang}, H., {Jiang}, X., {et~al.} 2016, PASP, 128, 115005

\bibitem[{Howell}(2000)]{Howell:2000}
{Howell}, S.~B. 2000, {Handbook of CCD Astronomy}

\bibitem[{Hu} {et~al.}(2014)]{Hu:2014RAA}
{Hu}, S.-M., {Han}, S.-H., {Guo}, D.-F., \& {Du}, J.-J. 2014, Research in
  Astronomy and Astrophysics, 14, 719

\bibitem[{Huang} {et~al.}(2012)]{Huang:2012RAA}
{Huang}, F., {Li}, J.-Z., {Wang}, X.-F., {et~al.} 2012, Research in Astronomy
  and Astrophysics, 12, 1585

\bibitem[{Jones}(2006)]{Howell:2006}
{Jones}, D. 2006, The Observatory, 126, 379

\bibitem[{Kinoshita} {et~al.}(2005)]{Kinoshita:2005ChJAA}
{Kinoshita}, D., {Chen}, C.-W., {Lin}, H.-C., {et~al.} 2005, CJAA, 5, 315

\bibitem[{Landolt}(1992)]{landolt:1992}
{Landolt}, A.~U. 1992, AJ, 104, 340

\bibitem[{Landolt}(2013)]{landolt:2013}
{Landolt}, A.~U. 2013, AJ, 146, 131

\bibitem[{Luo} {et~al.}(2012)]{Luo:2012}
{Luo}, Y.~P., {Zhang}, X.~B., {Deng}, L.~C., \& {Han}, Z.~W. 2012, APJL, 746,
  L7

\bibitem[{Shi} {et~al.}(1998)]{Shi:1998ChAA}
{Shi}, H.-M., {Qiao}, Q.-Y., {Hu}, J.-Y., \& {Lin}, Q. 1998, CHINESE ASTRONOMY
  AND ASTROPHYSICS, 22, 245

\bibitem[{Yang}(2013)]{YangYG:2013}
{Yang}, Y.-G. 2013, Research in Astronomy and Astrophysics, 13, 1471

\bibitem[{Zhang} {et~al.}(2016)]{Zhang:2016PASP}
{Zhang}, J.-C., {Fan}, Z., {Yan}, J.-Z., {et~al.} 2016, PASP, 128, 105004

\bibitem[Zhang {et~al.}(2015)]{ZhangJC:2015}
Zhang, J.-C., Ge, L., Lu, X.-M., {et~al.} 2015, Publications of the
  Astronomical Society of the Pacific, 127, 1292

\bibitem[Zhang \& Pi(2015)]{liyunzhang:2015}
Zhang, L., \& Pi, Q. 2015

\bibitem[{Zhang} {et~al.}(2015{\natexlab{a}})]{ZhangLY:2015}
{Zhang}, L., {Pi}, Q., {Han}, X.~L., {et~al.} 2015{\natexlab{a}}, NA, 38, 50

\bibitem[{Zhang} {et~al.}(2012)]{ZhangXB:2012}
{Zhang}, X.~B., {Deng}, L.~C., \& {Luo}, C.~Q. 2012, AJ, 144, 141

\bibitem[{Zhang} {et~al.}(2015{\natexlab{b}})]{ZhangYP:2015}
{Zhang}, Y.-P., {Jiang}, M.-D., {Zhang}, X.-B., {et~al.} 2015{\natexlab{b}},
  CHINESE ASTRONOMY AND ASTROPHYSICS, 39, 28

\bibitem[{Zhou} {et~al.}(2009)]{ZhouAY:2009}
{Zhou}, A.-Y., {Jiang}, X.-J., {Zhang}, Y.-P., \& {Wei}, J.-Y. 2009, Research
  in Astronomy and Astrophysics, 9, 349

\end{thebibliography}

\end{document}